\documentclass[letterpaper, 10 pt, conference]{ieeeconf}  

\IEEEoverridecommandlockouts                              
\usepackage{times}
\usepackage{epsfig}
\usepackage{graphicx}
\usepackage{amsmath}
\usepackage{amssymb}
\usepackage{mathptmx}
\usepackage{amsopn}
\usepackage{algorithm}
\usepackage{algorithmic}
\usepackage{multirow}
\usepackage{epstopdf}
\usepackage[caption=false]{subfig}
\usepackage[flushleft]{threeparttable}

\title{\LARGE \bf
Automatic Measurement of Physical Mobility in Get-Up-and-Go Test Using Kinect Sensor  
}

\author{Amir H. Kargar B., Ali Mollahosseini, Taylor Struemph \\ Wilson Pace, Rodney D. Nielsen and Mohammad H. Mahoor
\thanks{*This work is supported by the NSF grants IIS-1111568 and IIS-1262860.}
\thanks{{Amir H. Kargar B., Ali Mollahosseini, and Mohammad H. Mahoor are with the Department of Electrical and Computer Engineering at University of Denver, Denver, CO 80210. E-mails: {\tt\small \{akargarb, ali.mollahosseini, mmahoor\}@du.edu}}}
\thanks{ Taylor Struemph and Wilson Pace are with the Department of Family Medicine at University of Colorado, Aurora, CO 80045. E-mails: {\tt\small \{taylor.struemph, wilson.pace\}@ucdenver.edu}}
\thanks{Rodney D. Nielsen is with the Department of Computer Science and Engineering at University of North Texas, Denton, TX 76203. E-mail: {\tt\small rodney.nielsen@unt.edu}}
}

\begin{document}

\maketitle
\thispagestyle{empty}
\pagestyle{empty}

\begin{abstract}

Get-Up-and-Go Test is commonly used for assessing the physical mobility of the elderly by physicians. This paper presents a method for automatic analysis and classification of human gait in the Get-Up-and-Go Test using a Microsoft Kinect sensor. Two types of features are automatically extracted from the human skeleton data provided by the Kinect sensor. The first type of feature is related to the human gait (e.g., number of steps, step duration, and turning duration); whereas the other one describes the anatomical configuration (e.g., knee angles, leg angle, and distance between elbows). These features characterize the degree of human physical mobility. State-of-the-art machine learning algorithms (i.e. Bag of Words and Support Vector Machines) are used to classify the severity of gaits in 12 subjects with ages ranging between 65 and 90 enrolled in a pilot study. Our experimental results show that these features can discriminate between patients who have a high risk for falling and patients with a lower fall risk.

\end{abstract}  

\section{INTRODUCTION}
Human gait parameters are known to contain significant information about individuals' physical mobility and particularly mobility of the elderly who are at risk of falling. Clinicians have developed methods such as the Get-Up-and-Go Test (GUGT) ~\cite{TUG} for gait analysis and screening of the elderly. Such methods are able to predict subjects’ mobility and determine whether they can walk safely without the risk of falling. Since these assessments are done frequently, there is a great need for developing automatic and inexpensive computer systems capable of performing such assessments in the home of the elderly.

Different technical approaches and sensors have been proposed for gait assessment. One of the common approaches is the use of wearable sensors such as accelerometers or gyroscopes~\cite{accelerometers},~\cite{accelerometers1}. These devices mostly provide accurate gait parameters. Besides, not being expensive, having light weight and being small are other advantages of these sensors. However, these approaches would need a supervisor to help subjects wear the sensors and maintain them frequently. As a result, they are mostly suitable for laboratory purposes and not preferable for the elderly to use them frequently at home~\cite{demiris2004older}. 

On the other hand vision-based gait analysis systems have received great attention in recent years~\cite{poppe2007vision}. Some of these systems use regular RGB cameras while some others use more sophisticated sensors such as the Microsoft Kinect. In the camera-based systems, usually a calibrated array of cameras is utilized to provide a 3D representation of the scene. It has been shown that these methods are capable of providing an accurate model of the subject. For example in~\cite{leu2011robust}, two calibrated cameras were used to generate a 3D representation of the scene. Using the 3D model, various gait features were extracted including: torso angle, thigh angles and shank angles. However, the need of using more than one camera and their calibration and alignment issues have made it difficult to be used as an in-home screening system.

The Kinect sensor provides RGB images, depth range information and the human skeleton. Using the Kinect sensor for physical mobility assessment was first presented in~\cite{stone2011evaluation}. In that proposed method both a stereo-vision system and a Kinect sensor were used for comparison. In one experiment gait parameters were extracted using the Kinect depth output. In the second part, the same parameters were extracted using the output of the stereo vision system. Finally the results of both systems were compared with the results of a Vicon motion capture system. The experimental results showed that the Kinect sensor measures the gait parameters with a sufficient accuracy.

Another example is the work reported in~\cite{gabel2012full} where the skeleton data of a Kinect sensor was utilized for gait analysis. First some features were extracted from the skeleton data and were fed to a regression model. After that a state machine was used to produce desired states such as whether the foot touches the ground or not. In addition, other features such as the arm kinematics were measured, which show that a wide range of parameters can be extracted from the skeleton data. Nevertheless, they only extracted some features and no classification results were reported by the authors.

Using Kinect sensor eases the gait analysis and extraction of standard stride information with high accuracy in the home environment. Nonetheless, many researchers have focused on only extracting gait parameters for further analysis and there are limited works for automatic classification of subjects' degree of gait severity~\cite{accelerometers1}. Hence, designing an inexpensive system that extracts discriminative feature for accurate classification can be useful in alerting patients and clinicians without having a supervisor at home. This paper presents a methodology for classification of people into two categories, high fall-risk versus low fall-risk, based on their performance in the Get-Up-and-Go Test using Kinect sensor. In our approach, we first use image processing and computer vision algorithms to extract some desired features from the human skeleton data provided by a Kinect sensor. The features include number of steps, average step duration, and turn duration for gait parameters and distance between the elbows, angle between the legs, and angles between the shank and the thigh in each leg (knee angles) for anatomical parameters. Then using a Support Vector Machine (SVM) classifier, subjects are classified into those categories.

The rest of this paper is organized as follows. Section~\ref{GUGT} presents an overview of the Get-Up-and-Go Test and parameters that are considered as abnormalities in physical mobility. The proposed method using Kinect sensor for automatic measurement and classification of physical mobility is described in Section~\ref{processing}. The experimental results are presented in Section~\ref{result} and the paper is concluded in Section~\ref{sec:refrence}.

\section{Get-Up-and-Go Test}
\label{GUGT}
In terms of physical mobility a person is considered as independent if certain basic skills can be performed without any help of others~\cite{isaacs}. Get-Up-and-Go Test (GUGT) is a well-known simple test for mobility assessment which consists of basic everyday movements~\cite{GUGT}. In GUGT a subject sits on an arm chair, gets up, walks a three meter path, turns, walks back to the arm chair, and sits back down. In this test, subjects are asked to perform the task without any help from other people or objects (unless it is necessary), and physical mobility of the subject is rated on a scale of one to five according to the observation of a clinician. The problem with this method is the imprecision of the scoring system. A modified version of this test is called Timed Up and Go test~\cite{TUG}. This test computes a score based on the time taken by an individual to stand up from an arm chair, walk a distance of three meters, turn, walk back to the chair and sit down. Due to the timing, this version is more precise than the GUGT in scoring physical mobility. In~\cite{nordin}, it is shown that the time score and its variability between test trials correlates well with the physical mobility. 

Using the results of the GUGT and the Timed Up and Go test, clinicians can give an estimate of the overall muscle strength and balance of the body which can be used in fall prediction. Generally these tests show whether people are safe on their own or not when it comes to their mobility. At first, each subject sits on an arm chair and is ready to get up. One mobility indicator is the smoothness of getting up. When the subject uses any kind of help to get up or when getting up is not smooth, then it is a sign of abnormality. The next step is to start walking. Any gap between the time that the subject gets up and the time that he/she starts to walk is a sign of stabilizing and is abnormal. When the subject is walking, several abnormalities can be observed such as slow gait speed, feet dragging on the ground, deviating from the straight path, and having severe side to side movements. One significant part of the task is the turning part. Usually, one or two steps are enough to make a complete turn. Using more steps in turning, which corresponds to more turning time, is a sign of abnormality. In addition, some measurements on the anatomical configuration such as the angle between legs, the angles of knees and the distance between elbows, can be indicators of physical mobility while performing the test. In this paper, we extracted some of these abnormality indicators from a Kinect device automatically.




\section{AUTOMATIC PROCESSING OF GUGT}
\label{Sec:AUTOMATIC_PROCESSING}
\begin{figure}[thpb]
  \centering
  \includegraphics[scale=0.2]{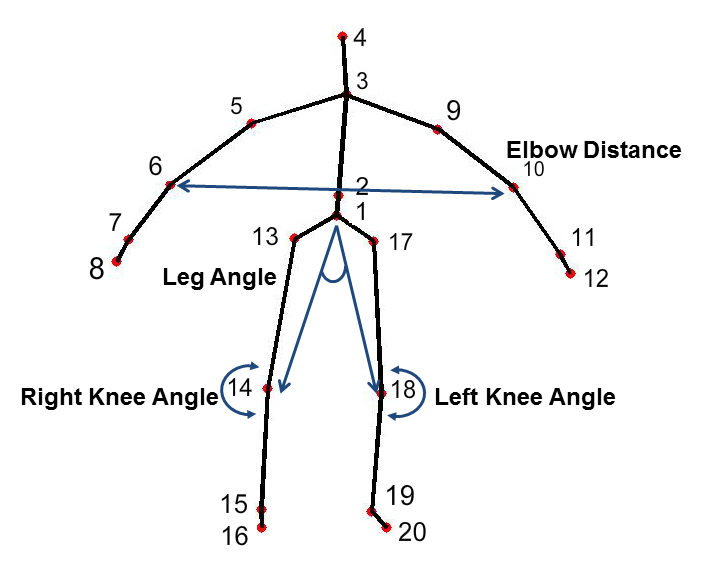}\\ 
  \vspace{-0.4cm}
  \caption{Layout of the Kinect skeleton data}
  \label{fig:joints}
\end{figure}

In order to measure the human's physical mobility in the GUGT, we used a Kinect sensor to capture a video and track the person's skeleton model while performing the test. The Microsoft Kinect sensor contains of an RGB camera, a depth sensor and a multi-array of microphones. The depth sensor consists of an infrared laser projector and an IR camera with the sensing range of 0.8 meters to 4.0 meters which captures depth images in resolution of 640$\times$480 pixels at 30 frames per second~\cite{microsoft}. It is capable of tracking the skeleton of one or two people moving within a practical range of 1.2 to 3.5 meters. The provided skeleton consists of 20 joints in the body with 30 frames per second. Fig.~\ref{fig:joints} shows a sample layout and joint indices of the virtual skeleton of the Kinect. 

\begin{figure}
\centering
\subfloat[][\label{Sitting}\scriptsize{Sitting}]
{
	\centering
    \includegraphics[scale = 0.16]{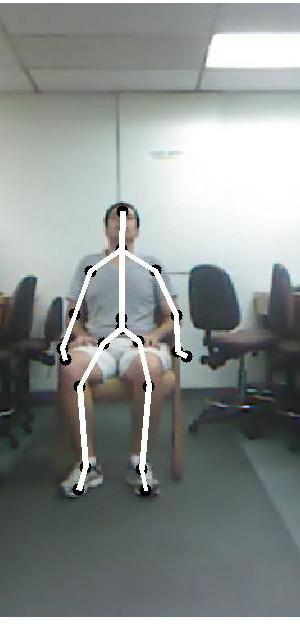}
}
\subfloat[][\label{Walking} \scriptsize{Walking}]
{
	\centering
    \includegraphics[scale = 0.16]{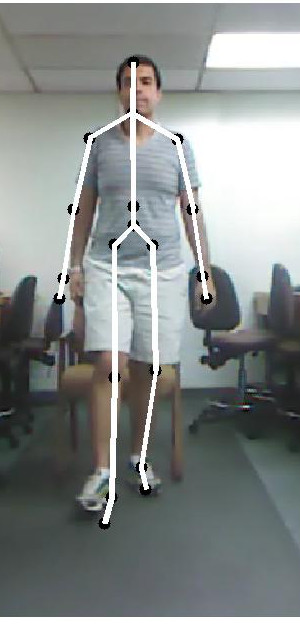}
}
\subfloat[][\label{Turning} \scriptsize{Turning}]
{
	\centering
    \includegraphics[scale = 0.16]{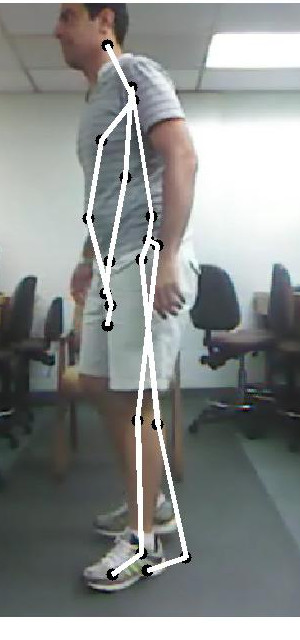}
}
\subfloat[][\label{Walking Back}\scriptsize{Walking-back}]
{
	\centering
    \includegraphics[scale = 0.16]{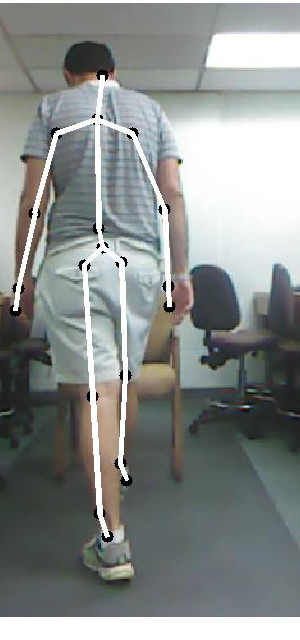}
}
\caption{ \label{fig:GUGTSamples} A subject performing the Get-Up-and-Go Test} 
\end{figure}

We collected a video dataset of the elderly people performing the GUGT while a Kinect sensor captured a video and tracked the person's skeleton model. Twelve subjects with ages ranging between 65 and 90 enrolled in the study. A geriatric physician reviewed the videos offline and categorized these subjects based on their gait movement and severity of their physical mobility into two categories. The first category includes patients that are relatively safe on their own with low fall risk, and the second one includes patients with high risk of falling that have severe physical mobility issues. The videos were then processed using image processing and computer vision algorithms aimed at extracting some features that were used for classification.

\label{processing}
\subsection{Feature Extraction}
\label{Sec:Feature_Extraction}
The Kinect sensor detects and tracks the positions ($x$, $y$ and $z$ ) of 20 joints of the human body skeleton. Since the joint points measured by Kinect can be noisy, we apply a median filter of size 5 to remove the noises and improve the accuracy of the measurements. Afterwards, two types of features are extracted: \emph{gait parameters} and \emph{anatomical parameters}. 
\subsubsection{Gait Parameters}
\label{Sec:Gait_Parameters}
In the GUGT a person is instructed to get up from an arm chair, walk, turn, walk back to the starting point, and sit back down. The path is automatically segmented into three phases including: \emph{Seated phase}, \emph{Walking phase}, and \emph{Turning phase}.

\begin{figure}
\centering
\subfloat[][\label{fig:Seatedphases}\scriptsize{The position of the hip joint in $z$-coordinate ($z_1$)}]
{
	\centering
    \includegraphics[scale = 0.32]{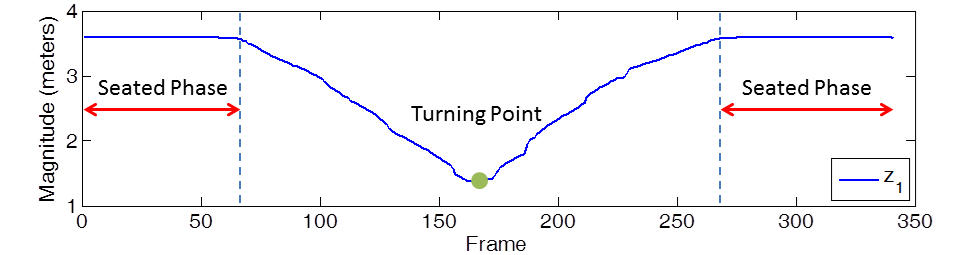}
}

\subfloat[][\label{fig:TuringPhase} \scriptsize{The absolute difference between $x$-coordinates of two elbows ($|x_6-x_{10}|$)}]
{
	\centering
    \includegraphics[scale = 0.32]{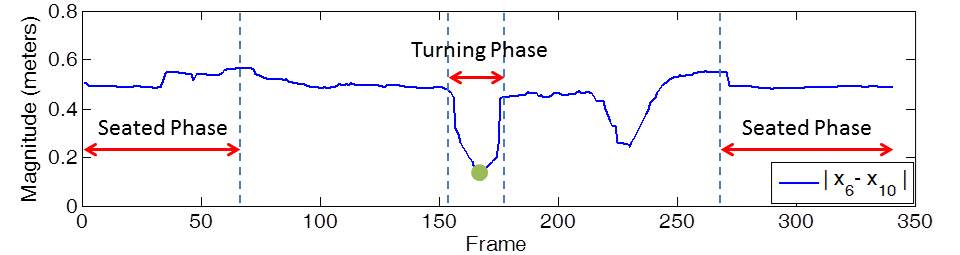}
}

\subfloat[][\label{fig:WalkingPhase} \scriptsize{The difference between $z$-coordinates of two heels ($z_{15}-z_{19}$)}]
{
	\centering
    \includegraphics[scale = 0.32]{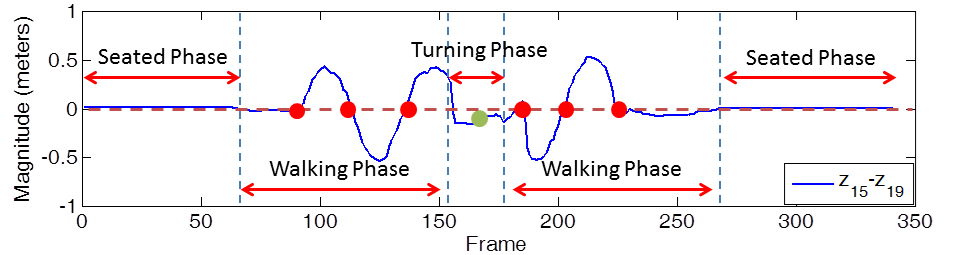}
}
\caption{The extracted seated, walking and turning phases in a GUGT}
\end{figure}

Using the position of the hip joint (point 1 in Fig.~\ref{fig:joints}), we can measure the distance of the person to the Kinect in $z$-direction. When the person is seated, the $z$-coordinate of the hip joint ($z_1$) does not change and is at its maximum. By measuring $z_1$ in consecutive frames from the beginning of the test, we can extract the seated phase. As the person starts walking towards the Kinect, $z_1$ decreases and when the person is turning, $z_1$ is at its minimum (we call this point as turn point). Fig.~\ref{fig:Seatedphases} shows the position of the hip joint in $z$-coordinate ($z_1$), the turn point, and the extracted seated phases. 

Since Kinect is mainly designed to track the human body when facing to the camera, it cannot recognize the skeleton of the body well while the person is turning (see Fig.~\ref{Turning}). The time that a person is turning can be defined as the time that the person starts to rotate his/her upper part of body to the time that rotation is done and the subject is ready to walk back. We used the absolute difference between $x$-coordinates of two elbows ($|x_6-x_{10}|$) to determine the starting and ending frames of a turn. In other word, when the person starts turning this difference decreases and when turning is finished, the absolute difference becomes the same as the value before turn. Measuring this distance in consequent frames before and after of the turn point, we can extract turning phase. Fig.~\ref{fig:TuringPhase} shows $|x_6-x_{10}|$ of a subject, turn point, and extracted turning phase. Extracting seated and turning phases, we can consider the frames in between as walking phases (Fig.~\ref{fig:WalkingPhase}).

Based on these three phases, three gait parameters are extracted. One of the gait parameters in walking phase is the number of steps that a person takes to perform GUGT. To detect steps, we used the difference between $z$-coordinates of two heels ($z_{15}-z_{19}$). Extremums of this difference indicate feet being far from each other and zeros correspond to feet being next to each other. Fig.~\ref{fig:WalkingPhase} shows the difference between $z$-coordinates of two heels and the extracted number of steps (where the difference is zero). 

Duration of each step is another feature which is important in physical mobility measurement. The difference between $z$-coordinates of two heels ($z_{15}-z_{19}$) gives us the starting frame and the ending frame of each step. For each skeleton data frame, Kinect provides a time-stamp which is used in calculating the duration of each step.

Number of steps in turning phase is another gait related property that contains significant information of the physical mobility. As it was explained before, Kinect cannot recognize the skeleton of the body while the person is turning. One approach to overcome this issue is to use another Kinect to look at the subject from side. This approach demands synchronization between Kinects which is against the simplicity of the proposed method. The other approach is to measure the turning time instead of counting the number of steps. We extracted the turning phase automatically and the turn duration is calculated based on the time-stamps of the starting and the ending frames of the turning phase.

\subsubsection{Anatomical Parameters}
The second type of parameters for physical mobility assessment is related to the anatomical configuration of the person while performing GUGT. Various anatomical measurements can be obtained to assess the condition of the body. One feature is the distance between two elbows (joints 6 and 10). The angle between the legs is another feature that is used here. This angle is defined as the angle between two vectors connecting joint 1 to 14 and 18. The other extracted features are the right and left knee angles that are the angles between the shank and the thigh in each leg. This angle is defined as the angle between two vectors connecting joint 18 to 17 and 19 for the left leg and the angle between two vectors connecting joint 14 to 13 and 15 for the right leg. These features are shown in Fig.~\ref{fig:joints}.


\subsection{Classification}
The features described in the previous section are used in classification of gait via a soft margin C-SVM classifier. The gait parameters provide us three numerical features for each sample which are: (1) The number of steps that have been taken to perform the test; (2) The average duration of steps in seconds; (3) The turn duration in seconds. These numbers can be fed into the classifier directly. 

The anatomical parameters can be easily measured in each frame of the skeleton data, but to be used as the input of the classifier, they should be comparable. In other words, there is no guarantee that subjects finish the task in the same amount of the time and hence there are different numbers of features for each subject in each test. To overcome this problem we have used the Bag-Of-Words (BOW)~\cite{fei2005bayesian}. The Bag-Of-Words or Bag-Of-Features is a simple approach that is commonly used in visual object classification, text categorization, etc., where there are different number of features for different samples. Assume there are \{$N_1,N_2,...,N_k$\} various number of features for $k$ samples, this model represents those features in $M$ keywords where in this paper $M \ll \{N_1,N_2,...,N_k\}$. For this purpose, first a clustering algorithm is used to cluster all training samples features to $M$ clusters. Each cluster is known by its center. Each feature of each sample is assigned to one of these centers, and then for each sample, the histogram of the features in these clusters represents new $M$ dimensional features of the sample. This $M$ dimensional output along with the three gait parameters are fed into the classifier. 


\section{Experimental Results}
\label{result}

The Kinect sensor was mounted on a table with an approximate height of 120cm. As the maximum practical range of Kinect sensor is 3.5 meters~\cite{microsoft}, and we want Kinect to view the whole body during the test, the arm chair was located at a distance of approximately 3.5 meters to Kinect and subjects walked a path of approximately two meters instead of three meters in original GUGT. Subjects, with their normal clothes, were instructed to first sit on the chair, stand up and start walking for two meters, then turn in place and walk back to the chair and sit back down. They were asked to walk completely normally during the test.


We measured the accuracy of the proposed classification technique on the dataset of 12 elderly patients collected at the clinic. Among those 12 patients, five subjects are categorized as low risk of falling and seven of them had high fall risk by an expert physician. Each patient repeated the test three to six times (each time is called a sample). A total of 50 samples were captured. We have used K-means clustering to cluster the anatomical parameters for the Bag-Of-Words and a C-SVM with radial basis kernel for classifying. K-means requires the number of clusters (K) to be defined which corresponds to the number of words in BOW. We evaluated the algorithm using 4 to 24 clusters, and it turned out that 10 clusters gave us the best classification rate.

To measure the performance of the proposed classification, we used a leave-one-subject-out technique. Particularly, the classifier is trained with samples of all subjects except one subject. For testing, the classifier classifies all samples of the left out subject. Every time one subject is left out and the training and testing procedure are repeated till all subjects are covered. Finally, the accuracy of the classification is reported as the average of samples being classified correctly. As K-means takes random initial clusters, BOW may generate different features each time which may affect the classification performance. We repeated the whole procedure ten times and on average the classification accuracy is 67.40\% with standard deviation of 4.72\%. Table \ref{classification_cMatrix} shows the average confusion matrix of the classification.

\begin{table}[h]
\caption{classification confusion matrix}
\vspace{-0.4cm}
\label{classification_cMatrix}
\centering
\begin{tabular}{l|c|c|}
\cline{2-3}
                                & Low Risk     & High Risk    \\ \hline
\multicolumn{1}{|l|}{Low Risk}  & \textbf{16.2} & 7.8           \\ \hline
\multicolumn{1}{|l|}{High Risk} & 8.5          & \textbf{17.5} \\ \hline
\end{tabular}
\end{table}

\section{Conclusion and Future Work}
\label{sec:refrence}
Results suggest that the classification method and features we designed provide an effective means of distinguishing patients who have a high risk for falling from patients with a lower fall risk. This inexpensive, easy to use, Kinect sensor-based approach can easily be used in the subject's home or by lower skilled healthcare personnel and relayed to a physician to further investigate as appropriate.

In this paper, we extracted only three gait parameters. Using depth data as well as skeleton tracking, we can extract more complicated gait parameters indicating abnormality in physical mobility such as smoothness of getting up, using any kind of help to get up, and feet dragging on the ground.

\addtolength{\textheight}{-12cm}   







\begin{thebibliography}{10}
\providecommand{\url}[1]{#1}
\csname url@samestyle\endcsname
\providecommand{\newblock}{\relax}
\providecommand{\bibinfo}[2]{#2}
\providecommand{\BIBentrySTDinterwordspacing}{\spaceskip=0pt\relax}
\providecommand{\BIBentryALTinterwordstretchfactor}{4}
\providecommand{\BIBentryALTinterwordspacing}{\spaceskip=\fontdimen2\font plus
\BIBentryALTinterwordstretchfactor\fontdimen3\font minus
  \fontdimen4\font\relax}
\providecommand{\BIBforeignlanguage}[2]{{%
\expandafter\ifx\csname l@#1\endcsname\relax
\typeout{** WARNING: IEEEtran.bst: No hyphenation pattern has been}%
\typeout{** loaded for the language `#1'. Using the pattern for}%
\typeout{** the default language instead.}%
\else
\language=\csname l@#1\endcsname
\fi
#2}}
\providecommand{\BIBdecl}{\relax}
\BIBdecl

\bibitem{TUG}
D.~Podsiadlo and S.~Richardson, ``The timed" up \& go": a test of basic
  functional mobility for frail elderly persons.'' \emph{Journal of the
  American geriatrics Society}, vol.~39, no.~2, pp. 142--148, 1991.

\bibitem{accelerometers}
K.~Culhane, M.~O'Connor, D.~Lyons, and G.~Lyons, ``Accelerometers in
  rehabilitation medicine for older adults,'' \emph{Age and ageing}, vol.~34,
  no.~6, pp. 556--560, 2005.

\bibitem{accelerometers1}
B.~R. Greene, A.~O. Donovan, R.~Romero-Ortuno, L.~Cogan, C.~N. Scanaill, and
  R.~A. Kenny, ``Quantitative falls risk assessment using the timed up and go
  test,'' \emph{Biomedical Engineering, IEEE Transactions on}, vol.~57, no.~12,
  pp. 2918--2926, 2010.

\bibitem{demiris2004older}
G.~Demiris, M.~J. Rantz, M.~A. Aud, K.~D. Marek, H.~W. Tyrer, M.~Skubic, and
  A.~A. Hussam, ``Older adults' attitudes towards and perceptions of'smart
  home'technologies: a pilot study,'' \emph{Informatics for Health and Social
  Care}, vol.~29, no.~2, pp. 87--94, 2004.

\bibitem{poppe2007vision}
R.~Poppe, ``Vision-based human motion analysis: An overview,'' \emph{Computer
  vision and image understanding}, vol. 108, no.~1, pp. 4--18, 2007.

\bibitem{leu2011robust}
A.~Leu, D.~Ristic-Durrant, and A.~Graser, ``A robust markerless vision-based
  human gait analysis system,'' in \emph{Applied Computational Intelligence and
  Informatics (SACI), 2011 6th IEEE International Symposium on}.\hskip 1em plus
  0.5em minus 0.4em\relax IEEE, 2011, pp. 415--420.

\bibitem{stone2011evaluation}
E.~E. Stone and M.~Skubic, ``Evaluation of an inexpensive depth camera for
  passive in-home fall risk assessment.'' in \emph{PervasiveHealth}, 2011, pp.
  71--77.

\bibitem{gabel2012full}
M.~Gabel, R.~Gilad-Bachrach, E.~Renshaw, and A.~Schuster, ``Full body gait
  analysis with kinect,'' in \emph{Engineering in Medicine and Biology Society
  (EMBC), 2012 Annual International Conference of the IEEE}.\hskip 1em plus
  0.5em minus 0.4em\relax IEEE, 2012, pp. 1964--1967.

\bibitem{isaacs}
B.~Isaacs, ``Clinical and laboratory studies of falls in old people. prospects
  for prevention.'' \emph{Clinics in geriatric medicine}, vol.~1, no.~3, pp.
  513--524, 1985.

\bibitem{GUGT}
S.~Mathias, U.~Nayak, B.~Isaacs \emph{et~al.}, ``Balance in elderly patients:
  the" get-up and go" test.'' \emph{Archives of physical medicine and
  rehabilitation}, vol.~67, no.~6, pp. 387--389, 1986.

\bibitem{nordin}
E.~Nordin, E.~Rosendahl, and L.~Lundin-Olsson, ``Timed up \& go test:
  reliability in older people dependent in activities of daily living?focus on
  cognitive state,'' \emph{Physical therapy}, vol.~86, no.~5, pp. 646--655,
  2006.

\bibitem{microsoft}
\BIBentryALTinterwordspacing
 [Online]. Available:
  \url{http://msdn.microsoft.com/en-us/library/hh973074.aspx}
\BIBentrySTDinterwordspacing

\bibitem{fei2005bayesian}
L.~Fei-Fei and P.~Perona, ``A bayesian hierarchical model for learning natural
  scene categories,'' in \emph{Computer Vision and Pattern Recognition, 2005.
  CVPR 2005. IEEE Computer Society Conference on}, vol.~2.\hskip 1em plus 0.5em
  minus 0.4em\relax IEEE, 2005, pp. 524--531.

\end{thebibliography}
\end{document}